\documentstyle[amssymb,aps,twocolumn,psfig]{revtex}

\begin{document}
\draft
\title{Generalizing the Extensive Postulates}
\author{L. Velazquez\thanks{%
luisberis@geo.upr.edu.cu}}
\address{Departamento de F\'{i}sica, Universidad de Pinar del R\'{i}o\\
Mart\'{i} 270, esq. 27 de Noviembre, Pinar del R\'{i}o, Cuba.}
\author{F. Guzm\'{a}n\thanks{%
guzman@info.isctn.edu.cu} }
\address{Departamento de F\'{i}sica Nuclear\\
Instituto Superior de Ciencias y Tecnolog\'{i}as Nucleares\\
Quinta de los Molinos. Ave Carlos III y Luaces, Plaza\\
Ciudad de La Habana, Cuba.}
\date{(Received: November 6, 2001)}
\maketitle

\begin{abstract}
We addressed the problem of generalizing the extensive postulates of the
standard thermodynamics in order to extend it to the study of nonextensive
systems. We did it in analogy with the traditional analysis, starting from
the microcanonical ensemble, but this time, considering its equivalence with
some generalized canonical ensemble in the thermodynamic limit by means of
the scaling properties of the fundamental physical observables.
\end{abstract}

\pacs{PACS: 05.20.Gg; 05.20.-y}

\section{Introduction}

Traditionally, the Statistical Mechanics and the standard Thermodynamics
have been formulated to be applied to the study of extensive systems, in
which the consideration of the additivity and homogeneity conditions, and
the realization of the thermodynamic limit are indispensable ingredients for
their good performance. In spite of its great success, this theory seems to
be inadequate to deal with many physical contexts. Nowadays it is known that
the above conditions\ can be considered as reasonable approximations for
those systems containing a huge number of particles interacting by means of
short range forces \cite{pat,stan,land} when these systems are not suffering
a phase transition \cite{Tiz,Gibb}.

In the last years considerable efforts have been devoted to the study of
nonextensive systems. The available and increasingly experimental evidences
on anomalies presented in the dynamical and macroscopic behavior in plasma
and turbulent fluids \cite{bog,bec,sol,shl}, astrophysical systems \cite
{lyn,pie,pos,kon,tor}, nuclear and atomic clusters \cite{gro,dago,ato},
granular matter \cite{kud} glasses \cite{par,stil} and complex systems \cite
{sta1} constitute a real motivation for the generalization of the
Thermodynamics.

One of the most important tendencies in the resolution of this problem is
the generalization of Boltzmann-Gibbs' formalism through the consideration
of a more general entropy form than the Shannon-Boltzmann-Gibbs':

\begin{equation}
S_{SBG}=-%
\mathrel{\mathop{\sum }\limits_{k}}%
p_{k}\ln p_{k}.
\end{equation}
This is the case of the Renyi \cite{renyi} and Sharma-Mittal \cite{shamit}
entropies and the so popular Tsallis non-extensive entropy \cite{tsal}, to
mention some of them. It has been said that in the second half of last
century close to 30 different entropic forms have been proposed! This number
is increasing continuously through the years.

The main advantage of all these Probabilistic Thermodynamic Formalisms
(PThF), those based on the consideration of the concept of information
entropy, is the versatility of applications, be already well inside the
physical sceneries, in the study of the behavior of systems during the
equilibrium or dynamical processes, as well as other interdisciplinary
branches of knowledge. However, in our opinion, this kind of theory
undergoes two remarkable defects. The first, the non-uniqueness of \
information entropy concept, which is evident. The second, it is not
possible to determine the application domain {\it a priori} for each
specific entropic form. For example, in the case of the Tsallis'
nonextensive statistics we can refer the determination of the entropic index 
$q$. That is why we consider that the statistical description of
nonextensive systems must start from their microscopic characteristics.

Ordinarily, the macroscopic description of systems in thermodynamic
equilibrium through its microscopic properties has the starting point in
considerating the microcanonical ensemble. Although there is no way to
justify this selection, it is generally accepted its great significance
since from this ensemble the Thermostatistics could be deduced without
invoking anything outside the mechanics.

Boltzmann \cite{Bolt}, Gibbs \cite{Gibb}, Einstein \cite{Ein1,Ein2} and
Ehrenfests \cite{Ehre} recognized the hierarchical primacy of the
microcanonical ensemble with respect to the canonical and Gran-canonical
ensembles. The last ones can be derived from the first by considerating
certain particular conditions: {\it the extensivity postulates}. These are
the essential ideas to generalize the traditional results for the
justification of any generalized Boltzmann-Gibbs' formalism, to put these on
the level of the microscopic description, in the ground of Mechanics.

The present paper will be devoted to the analysis of the necessary
conditions to be taken into account in order to perform this derivation. The
same will be done in analogy with the traditional case, starting from the
microcanonical ensemble, but this time generalizing the traditional
postulates to cover systems with long-range interactions.

\section{Microcanonical ensemble: its geometric aspects and scaling
properties.}

Let us show now that a reasonable way to generalize the traditional
Thermodynamic is assuming the {\bf scaling postulates}: The {\bf equivalence}
of the microcanonical description with a generalized canonical one during
the realization of the {\bf Thermodynamic Limit} by means of the {\bf %
self-similarity scaling properties} of the fundamental physical observables.

\subsection{The thermodynamic limit.}

Analyzing the three postulates that sustain the validity of the standard
thermodynamic it is easy to see that the realization of the thermodynamic
limit possesses a most universal character. In fact, the additivity and
homogeneity conditions are associated more indissolubly to the short-range
nature of the interactions in comparison with the characteristic linear
dimension of a sufficient large system.

However, in the real world we can find examples of systems where the
thermodynamic limit does not take place, since they are not composed by a
huge number of particles. As examples of these, it can be mentioned the
molecular and atomic clusters. A very interesting problem is the extension
of the macroscopic description for this kind of systems as well as the
question about when it could be considered that these systems have reached
the thermodynamic limit.

An important step toward the resolution of these problems was accomplished
by D.H.E. Gross with the development of the {\it Microcanonical
Thermostatistics }\cite{gro,gro2,gro3}, a formalism based on the
consideration of the microcanonical ensemble. In this approach is possible
to accomplish the description of some finite systems, extending in this way
the study of phase transitions to them. On the other hand, its possibilities
are very attractive when exploring the behavior of the macroscopic
observables with the increase of the number of particles and analyzing the
convergence in some cases in an ordinary extensive system, allowing us in
this way to give a valuation about when the thermodynamic limit is
established in the system.

A remarkable aspect of this approach is that the justification of the
canonical ensemble is not supported by the consideration of the Gibbs'
argument:{\em \ }a closed system composed by a subsystem with a weak
interaction with a thermal bath,{\em \ but by means of equivalence between
the microcanonical and canonical ensemble in the thermodynamic limit for
this kind of system}. The Gibbs' argument is sustained by the possibility of
the separation of this subsystem from the whole system, that is,\ in the
independency or weak correlation of this subsystem with the remaining part.
It is easy to understand that this argument can not be applied for many
nonextensive systems, since in them there are long-range correlations due to
the presence of long-range interactions.

Thus, {\em the equivalence of the microcanonical ensemble with the canonical
ensemble during the thermodynamic limit could be easily extended for the
nonextensive systems since this argument has an universal character stronger
than the Gibbs'}. We have to point out that this equivalence also supports
the thermodynamic formalism based on the Legendre Transformation between the
thermodynamic potentials, which is usually valid without restriction for the
probabilistic interpretation of the thermodynamics \cite{gro2}.

The thermodynamic formalism of the Gross' theory have been defined in a way
that be equivalent in the thermodynamic limit with the ordinary form of the
Thermodynamic for the case of the extensive systems. That is why this theory
is appropriate to study systems that in the thermodynamic limit become in
ordinary extensive system\ (see for example in ref. \cite{gro4,gro5}).
However it is not trivial its application to systems with long-range
interactions because we do not know their asymptotical properties as well as
the corresponding generalized canonical ensemble.

\subsection{Geometrical aspects of the PDF.}

A feature of the\ PThF is that the description of the macroscopic state of
the systems is performed by means of the intensive parameters of the
generalized canonical distributions, $\beta $, its means, they are supported
by the validity of some {\it generalized\ zero principle of the
Thermodynamics}. This section will be devoted to the analysis of the
prescriptions for this kind of description.

Let us pay special attention to the geometrical aspects that posses the
Probabilistic Distribution Functions (PDF) of the ensembles. For the case of
the microcanonical ensemble its PDF is given by:

\begin{equation}
\omega _{M}\left( X;I,N,a\right) =\frac{1}{\Omega \left( I,N,a\right) }%
\delta \left[ I-I_{N}\left( X;a\right) \right] \text{,}
\end{equation}
where ${\cal R}_{I}\equiv \left\{ I_{N}\left( X;a\right) \right\} $ is the
set of movement integrals of the distribution, and $\Omega \left(
I,N,a\right) $ is the accesible states density of the microscopic state. Any
function of the above movement integrals is itself a movement integral. If
we have a complete independent set of such a functions this set of functions
will be equivalent to the set ${\cal R}_{I}$, that is, both of them
represent the same macroscopic state. That is why it is more exactly to
speak about the movement integrals abstract space of the microcanonical
distribution, $\Im _{N}$. Therefore, every physical quantity or behavior has
to be equally reproduced by any representation of the space of macroscopic
state, $\left( \Im _{N};a\right) $.

It is easy to show that the microcanonical distribution is invariant under
the transformation group of {\it local reparametrizations or Diffeomorfism}
of the movement integrals space, $Diff\left( \Im _{N}\right) .$ Let $\Re
_{I} $ and $\Re _{\phi }$ be two representations of $\Im _{N}$, where: 
\begin{equation}
\ I\equiv \left\{ I^{1},I^{2}\ldots I^{n}\right\} \ \text{and}\ \phi \equiv
\left\{ \phi ^{1}\left( I\right) ,\phi ^{2}\left( I\right) \ldots \phi
^{n}\left( I\right) \right\} \text{.}
\end{equation}
\ \ By the property of the $\delta $-function we have:

\begin{equation}
\delta \left[ \phi \left( I\right) -\phi \left( I_{N}\left( X;a\right)
\right) \right] =\left| \frac{\partial \phi }{\partial I}\right| ^{-1}\delta %
\left[ I-I_{N}\left( X;a\right) \right] \text{,}
\end{equation}
hence:

\begin{equation}
\widetilde{\Omega }\left( \phi ,N,a\right) =\left| \frac{\partial \phi }{%
\partial I}\right| ^{-1}\Omega \left( I,N,a\right) \text{,}
\end{equation}
and therefore:

\begin{equation}
\omega _{M}\left( X;\phi ,N,a\right) =\omega _{M}\left( X;I,N,a\right)
\equiv \omega _{M}\left( X;\Im _{N},a\right) \text{.}
\end{equation}

From here, two interesting remarks are derived. {\bf The first}: This local
reparametrization invariance is the maximal symmetry that a geometrical
theory could possess, and it is usually associated to the local properties
of a curved general space. This property is in complete contradiction with
the{\em \ ergodic chauvinism} of the traditional distributions. This is
translated as the preference of the {\em energy} over all the integrals of
movement corresponding to the system. In the microcanonical ensemble this
preference is not justified since it does not depend on the representation
of the movement integrals space. {\bf The second}: The movement integrals
are defined by the commutativity relation with the Hamiltonian of the
system. In the case of closed systems, the hamiltonian is the total energy,
a conserved quantity. In the microcanonical distribution it represents one
of the coordinates of the point belonging to $\Im _{N}$, in an specific
representation. When we change the representation, the energy loses its
identity. Since we can not fix the commutativity of the energy with the
other integrals, we must impose that all movement integrals commute between
them in order to preserve these conditions. As we see, even from the
classical point of view, {\em the reparametrization invariance suggests that
the set of movement integrals have to be simultaneously measurable}. In the
quantum case, this is an indispensable request for the correct definition of
the statistical distribution. This property is landed to the classical
distribution by the correspondence between the Quantum and the Classic
Physics.

On the other hand, all the PDF derived from the PThF must depend on the
movement integrals in a lineal combination with the canonical intensive
parameters: 
\begin{equation}
p\left( X;\beta ,N\right) =F(\beta ,N;\beta \cdot I_{N}\left( X\right) )%
\text{,}
\end{equation}
therefore, the most general symmetry group that preserves this functional
form is the {\em local unitary lineal group}, $SL_{c}\left( {\bf R}^{{\bf n}%
}\right) $ ($n$ is the dimension of $\Im _{N}$), which is related with the
euclidean vectorial spaces. It is supposed that an {\em spontaneous symmetry
breaking} takes place during the thermodynamic limit, from $Diff\left( \Im
_{N}\right) $ to $SL_{c}\left( \text{{\bf \ss }}^{{\bf n}}\right) $, where 
{\bf \ss }$^{{\bf n}}$ is the space of the generalized canonical parameters.
The way in which this hypothetical symmetry breaking happens will determine
the specific form of the PDF in the thermodynamic limit.

What do we understand as an spontaneous symmetry breaking? Ordinarily, the
microcanonical ensemble could be described equivalently by any
representation of the movement integrals space. However, with the increasing
of the system degrees of freedom, some specific representations will be more
adequate to describe the macroscopic state because it shows in a better way
the general properties of the system ( i.e., in the case of the traditional
systems, {\it the extensivity)}.

\subsection{The scaling symmetry of the fundamental physical observables.}

In complete analogy with the traditional analysis, we identify the cause of
this spontaneous symmetry breaking with the {\em scaling properties of the
fundamental macroscopic observables during the thermodynamic limit}: the
behavior of the movement integrals, the external parameters and the
accessible states density with the increasing of the system degrees of
freedom.

As it was pointed out previously, these symmetry properties will be just
expressed for certain set of representations, ${\cal M=}\left\{ \Re
_{I^{\ast }}\right\} $. Only in the frame of this set of admissible
representations will be valid the equivalence between the microcanonical
description and the generalized canonical one. The remaining symmetry among
representations belonging to ${\cal M}$ constitutes the local group $%
SL_{c}\left( {\bf R}^{{\bf n}}\right) $ of the canonical description.

In order to access to the scaling behavior of the movement integrals, the
asymptotical dependence of the states density for large values of the number
of particles, $W_{asym}\left( I^{\ast },N,a\right) $, must be obtained:

\begin{equation}
W\left( I^{\ast },N,a\right) 
\mathrel{\mathop{\Rightarrow }\limits_{N\gg 1}}%
W_{asym}\left( I^{\ast },N,a\right) ,
\end{equation}
where $\Re _{I^{\ast }}$ $\in {\cal M}$. Let $W_{o}$ be defined as:

\begin{equation}
W_{o}=W_{asym}\left( I_{o}^{\ast },N_{o},a_{o}\right) ,
\end{equation}
where $N_{o}$ could be considered to be in the asymptotical region $N\gg 1$.
The scaling behavior of the macroscopic observables is dictated by the
symmetry of the asymptotical states density under the{\em \ }{\bf %
self-similar scaling transformations}:

\begin{equation}
\left. 
\begin{array}{c}
N_{o}\rightarrow N=\alpha N_{o} \\ 
I_{o}^{\ast }\rightarrow I^{\ast }=\alpha ^{\varkappa }I_{o}^{\ast } \\ 
a_{o}\rightarrow a=\alpha ^{\pi _{a}}a_{o}
\end{array}
\right\} \Rightarrow W_{asym}\left( I^{\ast },N,a\right) ={\cal F}\left[
W_{o},\alpha \right] \text{.}
\end{equation}

In the above definition, $\alpha $ is the scaling parameter, $\varkappa $
and $\pi _{a}$ are real constants characterizing the scaling
transformations. As we see, all the movement integrals are transformed by
the same scaling law in order to satisfy the $SL\left( {\bf R}^{n}\right) $
invariance of the PDF in the thermodynamic limit. Finally, ${\cal F}$ is a
functional dependence of $W_{o}$ and $\alpha $ satisfying the
self-similarity condition: 
\begin{equation}
\text{2.- }{\cal F}\left[ {\cal F}\left[ x,\alpha _{1}\right] ,\alpha _{2}%
\right] ={\cal F}\left[ x,\alpha _{1}\alpha _{2}\right] \text{.}
\end{equation}
The above request could be satisfied by the following generic form:

\begin{equation}
\text{ }{\cal F}\left[ x,\alpha \right] =f_{\infty }\left[ \alpha ^{\sigma
}f_{\infty }^{-1}\left( x\right) \right] \text{,}
\end{equation}
where $f_{\infty }\left( x\right) $ is an increasing{\bf \ }function of $x$
with a monotonic first derivative in the asymptotical region $x\gg 1$, and $%
\sigma $, a real constant. The function $f_{\infty }\left( x\right) $
defines the specific {\bf scaling law} of the system. Let us introduce the
function $f\left( x\right) $, which is an increasing function of $x$ with a
monotonic first derivative for all the positive real values of $x$
satisfying the following conditions:

\begin{equation}
\text{\ }f\left( 0\right) =1\text{ (Normalization),}
\end{equation}

\begin{equation}
\ f\left( x\right) 
\mathrel{\mathop{\rightarrow }\limits_{x\gg 1}}%
f_{\infty }\left( x\right) \text{ (Convergency).}
\end{equation}

This function allows us to define the function $\gamma \left( I,N;a\right) $:

\begin{equation}
\gamma \left( I^{\ast },N;a\right) =\frac{1}{N^{\sigma }}f^{-1}\left[
W\left( I^{\ast },N;a\right) \right] \text{,}
\end{equation}
which is scaling invariant during the realization of the thermodynamic
limit. It is not difficult to understand that the objective of the canonical
description is the knowledge of the set of admissible representations, $%
{\cal M}$, the scaling laws of the fundamental macroscopic observables
(integrals of movement and external parameters) as well as the determination
of the scaling invariant function $\gamma _{\infty }\left( I^{\ast
},N;a\right) $:

\begin{equation}
\gamma _{\infty }\left( I^{\ast },N;a\right) =%
\mathrel{\mathop{\lim }\limits_{\alpha \rightarrow \infty }}%
\gamma \left( I^{\ast },N;a\right) \equiv \frac{1}{N_{o}^{\sigma }}f_{\infty
}^{-1}\left[ W\left( I_{o}^{\ast },N_{o};a_{o}\right) \right] \text{.}
\end{equation}

Although $\gamma _{\infty }\left( I^{\ast },N;a\right) $ is not a physical
observable, from this function could be derived other quantities which are
experimentally mensurable, allowing in this way to obtain information about
the ordering of the system. This fact aims at the following conclusion: {\em %
In the generalized canonical description, the function}:

\begin{equation}
S_{f}=f^{-1}\left[ W\left( I^{\ast },N;a\right) \right] \text{,}  \label{gbp}
\end{equation}
{\em plays the role of the }{\bf generalized entropy}. From now on we will
refer the definition of the Eq.(\ref{gbp}) as the {\bf generalized Boltzmann%
\'{}%
s Principle}. The function $f^{-1}\left( x\right) $ defines the so called 
{\em counting rule for the entropy}. Our proposition suggests the relative
significance of the entropy concept. In the thermodynamic limit the
generalized Boltzmann%
\'{}%
s entropy may possess a probabilistic interpretation with the same style of
the Shannon-Boltzmann-Gibbs or the Tsallis nonextensive entropy. The
specific form of the thermodynamic formalism could be determined by means of
the conditions for the equivalence of the ensembles, maybe, through some
generalization of the ordinary Laplace's Transformation \cite{gro2}.

A final remark: Ordinarily, the validity of the equivalence between the
microcanonical and the canonical ensemble must satisfy a stability
condition, which is ordinarily improved by the exigency of the concavity of
the entropy function. This mathematical demand is intimately bounded to the
homogeneity condition, the non occurrence of phase transitions in the
system. In the case of the finite system this exigency leads to the
generalization of the microcanonical thermostatistic of D.H.E. Gross.

\section{Conclusions}

We have analyzed a possible way to generalize the extensive postulates in
order to extend the Thermodynamics to the study of some nonextensive
systems. We showed the superiority of the argument of the equivalence of the
microcanonical ensemble with some generalized canonical ensemble with regard
to the Gibbs' one. The ergodic chauvinism of the classical thermodynamics is
not justified in the microcanonical ensemble since this description is
reparametrization invariant. This property has a relevant role in the
passage from one description to the another, with the occurrence of a
hypothetical spontaneous breaking of this symmetry in the thermodynamic
limit by means of the scaling properties of the fundamental physical
observables. The generalized Boltzmann%
\'{}%
s Principle, the definition of the generalized Boltzmann%
\'{}%
s entropy [the Eq.(\ref{gbp})], have been defined in order to deal the study
of systems with an arbitrary scaling laws [ i.e., {\em exponentials}, {\em %
potentials (fractals)}, etc.].

The previous arguments can be applied to the study of some nonextensive
systems, allowing in this way to justify a thermodynamic probabilistic
formalism analogue to the traditional one without the necessity of appealing
to additional postulates.

\newpage
\onecolumn

\end{document}